# An Exact Linearization Method for OLTC of Transformer in Branch Flow Model

Wenchuan Wu, *Senior Member, IEEE*, Zhuang Tian, Boming Zhang, *Fellow, IEEE*

*Abstract*--The branch flow based optimal power flow (OPF) problem in radially operated distribution networks can be exactly relaxed to a second order cone programming (SOCP) model without considering transformers. However, the introduction of nonlinear transformer models will make the OPF model non-convex. This letter presents an exact linearized transformer's OLTC (On-Load Tap-Changer) model to keep the OPF model convex via binary expansion scheme and big-M method. Validity of the proposed method is verified using IEEE 33-bus test network.

[1]*Index Terms*—distribution networks, optimal power flow, OLTC

## I. INTRODUCTION

OPTIMAL power flow is an essential method to reduce network losses and optimize voltage profile in distribution networks. In [1], the relaxation and convexification of branch flow based OPF model are well discussed, which is ultimately formulated as a second-order cone programming (SOCP) problem (SOCP-OPF). However, the transformer has been ignored in this model, which is critical to improve voltage profile. In published works, transformer tap ratio is usually approximately modeled [2] or relaxed to continuous variables [3], which may lead to suboptimal or infeasible solutions [2]. In this letter, an exact linearized transformer's OLTC model is proposed and incorporated into the SOCP-OPF model.

## II. LINEARIZATION OF TRANSFORMER'S OLTC MODEL

Fig. 1 shows a partial distribution feeder with a transformer.

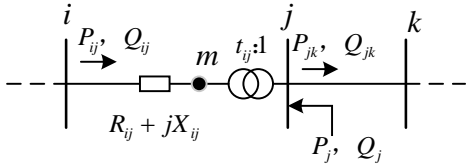

Fig. 1 One line diagram of a feeder with a transformer

Transformer *ij* can be divided into branch *im* and branch *mj*, branch *mj* only contains a tap-changer, and the impedance of branch *im* is the same as branch *ij*. Similar with Distflow branch equations for lines [4], the branch equations of transformer *ij* can be described as

$$P_{ij} - L_{ij}R_{ij} + P_j = P_{jk} \quad (1)$$
$$Q_{ij} - L_{ij}X_{ij} + Q_j = Q_{jk} \quad (2)$$

W. Wu, Z. Tian, B. Zhang are with the State Key Laboratory of Power Systems, Department of Electrical Engineering, Tsinghua University, Beijing 100084, China (e-mail: wuwench@tsinghua.edu.cn).

$$U_i - U_{jt} = 2(R_{ij}P_{ij} + X_{ij}Q_{ij}) - (R_{ij}^2 + X_{ij}^2)L_{ij} \quad (3)$$
$$L_{ij} = (P_{ij}^2 + Q_{ij}^2)/U_i \quad (4)$$
$$U_{jt} = t_{ij}^2 U_j \quad (5)$$

where $P_{ij}$ and $Q_{ij}$ represent the sending-end active and reactive power from buses $i$ to $j$; $P_{jk}$ and $Q_{jk}$ represent the sending-end active and reactive power from buses $j$ to $k$; $U_i$, $U_j$ and $U_{jt}$ represent the square of voltage magnitude of buses $i$, $j$ and $m$; $L_{ij}$ represent the square of current magnitude of branch $ij$; $P_j$ and $Q_j$ represent the active and reactive power injection at bus $j$; $R_{ij}$ and $X_{ij}$ represent the impedance on branch $ij$; $t_{ij}$ represents the turns ratio of the transformer in branch $ij$.

The branch flow model (1)-(5) is non-convex due to equations (4) and (5). Equation (4) can be relaxed to a convex constraint by SOCP relaxation [1]. Equation (5) can be transformed into mixed integer linear expressions by the method proposed in this letter.

The transformer's tap-changer is formulated as follows:
$$t_{ij} = t_{ij}^{min} + T_{ij}\Delta t_{ij}, 0 \leq T_{ij} \leq K_{ij} \quad (6)$$
$$\Delta t_{ij} = (t_{ij}^{max} - t_{ij}^{min})/K_{ij} \quad (7)$$

where $t_{ij}^{min}$ and $t_{ij}^{max}$ represent the minimum and maximum turns ratios of the transformer; $K_{ij}$ represents the total taps of the tap-changer; $\Delta t_{ij}$ represents the turns ratio change per tap; $T_{ij}$ is an integer variable which represents the actual tap position of the tap-changer.

$t_{ij}$ can be expressed by the binary expansion scheme [5] as:
$$t_{ij} = t_{ij}^{min} + \Delta t_{ij} \sum_{n=0}^{N_{ij}} 2^n \lambda_{ij,n} \quad (8)$$
$$\sum_{n=0}^{N_{ij}} 2^n \lambda_{ij,n} \leq K_{ij} \quad (9)$$

where $\lambda_{ij,n}$ is a binary variable and $N_{ij}$ is the length of the binary representation of $K_{ij}$.

Multiplying both sides of (8) by $U_j$ and defining new variables $m_{ij} = t_{ij}U_j$ and $x_{ij,n} = \lambda_{ij,n}U_j$, we obtain

$$m_{ij} = t_{ij}^{min}U_j + \Delta t_{ij}\sum_{n=0}^{N_{ij}} 2^n x_{ij,n} \quad (10)$$

$x_{ij,n} = \lambda_{ij,n}U_j$ can be equivalently replaced by equations (11) and (12) by introducing a positive large number $M$:

$$0 \leq U_j - x_{ij,n} \leq (1-\lambda_{ij,n})M \quad (11)$$

$$0 \leq x_{ij,n} \leq \lambda_{ij,n}M \quad (12)$$

Applying the same procedure to $U_{jt} = t_{ij}m_{ij}$ and defining a new variable $y_{ij,n} = \lambda_{ij,n}m_{ij}$, we obtain

$$U_{jt} = t_{ij}^{\min}m_{ij} + \Delta t_{ij}\sum_{n=0}^{N_{ij}} 2^n y_{ij,n} \quad (13)$$

$$0 \leq m_{ij} - y_{ij,n} \leq (1-\lambda_{ij,n})M \quad (14)$$

$$0 \leq y_{ij,n} \leq \lambda_{ij,n}M \quad (15)$$

With the above transformations, the non-convex constraints of transformer (5)-(7) are transformed into mixed integer linear constraints (9), (10)-(15).

In [2], binary expansion scheme has been used to linearize an approximate transformer's OLTC model as follows:

$$U_{jt} = (t_{ij}^{\min})^2 U_j + 2T_{ij}\Delta t_{ij}U_j \quad (16)$$

Equation (16) is inaccurate and the optimized turns ratios deviate from their exact values. This method will be compared with our method in section III.

The linearized transformer model can be incorporated into the SOCP-OPF [1]. The objective of OPF is to minimize active power losses

$$\min \sum_{i \in \Phi_{all}} P_i \quad (17)$$

subject to

1) Active and reactive power balance constraints
2) Constraints for line branches

$$U_i - U_j = 2(R_{ij}P_{ij} + X_{ij}Q_{ij}) - (R_{ij}^2 + X_{ij}^2)L_{ij}, \forall ij \in \Phi_l \setminus \Phi_{ltf} \quad (18)$$

3) Constraints for transformer branches

$$(3), (7), (9), (10)–(15), \forall ij \in \Phi_{ltf} \quad (19)$$

4) Second-order cone constraints

$$L_{ij} \geq (P_{ij}^2 + Q_{ij}^2)/U_i, \forall ij \in \Phi_l \quad (20)$$

5) Constrains for power generation and consumption
6) Constrains for voltage and current magnitudes

where $\Phi_{all}$ represents the set of all buses; $K(j)$ represents the set of buses connected to bus $j$; $\Phi_l$ represents the set of all branches; $\Phi_{ltf}$ represents the set of transformer branches.

The above model is a mixed integer SOCP problem and can be solved using software packages such as CPLEX.

## III. NUMERICAL EXAMPLE

The proposed OPF problem is tested on a modified IEEE 33-bus system [4]. Transformers are installed on branch 1-2, 2-19, 3-23 and 6-26. The turns ratio of each transformer varies between 0.95 to 1.05 with same change per tap, denoted by $\Delta t$. DGs are installed at buses 6, 14 and 29. The minimum and maximum allowable bus voltages are 0.95 p.u. and 1.05 p.u., respectively.

The program is implemented using YALMIP® and CPLEX®. Numerical tests are performed on a computer with Intel® Core™ i5 (2.60 GHz) and 12 GB RAM. In order to illustrate the effectiveness of the transformer linearization method and extra computation time caused by the mixed integer linear constraints, six scenarios are tested: 1) $\Delta t = 0.02$; 2) $\Delta t = 0.01$; 3) $\Delta t = 0.005$; 4) $\Delta t = 0.002$; 5) $\Delta t = 0.001$; 6) turns ratios fixed to the results of scenario 3. The power losses and computation time of our method and the method in [6] for each scenario are listed in Table I.

Table I Simulation results for 33-bus system

| Scenario | Taps | Power Losses (kW) | CPU Time (s) | |
|---|---|---|---|---|
| | | | Our method | Method in [6] |
| 1 | 5 | 41.2 | 1.03 | 1.40 |
| 2 | 10 | 41.0 | 1.60 | 2.01 |
| 3 | 20 | 41.0 | 1.62 | 2.17 |
| 4 | 50 | 40.9 | 2.37 | 5.17 |
| 5 | 100 | 40.9 | 2.72 | 14.60 |
| 6 | 200 | 40.9 | 2.84 | 39.19 |

From the results of scenarios 1, 2 and 3, we can see that more power losses can be reduced by optimizing tap-changers with more taps, while the computation times grow due to the increasing of binary variables. However, the additional computation burden introduced by the transformer constraints is moderate (comparing scenario 3 with 6) because the binary expansion scheme does not involve large number of binary variables in transformer model. And the proposed method has higher computational efficiency compared to [6] due to less binary variables.

The approximate method in [2] and the proposed exact method in this paper are tested with $\Delta t = 0.005$. The turns ratios and power losses are listed in Table II.

Table II Results of our method and method in [2]

| Method | $t_{2-1}$ | $t_{19-2}$ | $t_{23-3}$ | $t_{26-6}$ | Power Losses (kW) |
|---|---|---|---|---|---|
| Method in [2] | 1.05 | 1.0012 | 1.0012 | 1.0062 | 41.1 |
| Our method | 1.05 | 1 | 1.005 | 1.01 | 41.0 |

From Table II, we can see that the turns ratios solved by the approximate method [2] are not accurate. The method proposed in this paper is exact and the solution is optimal.

## IV. CONCLUSIONS

The nonlinear transformer model is exactly linearized by binary expansion scheme and big-M method. This exact linearized transformer model involves few binary variables and can be incorporated into the SOCP-OPF problem in distribution networks. The proposed method has been simulated on a IEEE 33-bus system under different scenarios to demonstrate its effectiveness.